\documentclass[aps,prb,twocolumn,groupedaddress,showpacs]{revtex4}
\usepackage{color}
\usepackage{epsfig}

\begin{document}
 
\title{Monte Carlo Study of an Inhomogeneous Blume-Capel Model}
\author{S.M. Pittman$^1$, G.G. Batrouni$^2$, and
  R.T. Scalettar$^1$}
\affiliation{$^1$Physics Department, University of California, 
Davis, California 95616, USA}
\affiliation{$^2$INLN, Universit\'e de Nice-Sophia Antipolis, CNRS; 
1361 route des Lucioles, 06560 Valbonne, France}

\begin{abstract}
Systems of particles in a confining potential exhibit a spatially
dependent density which fundamentally alters the nature of phase
transitions that occur.  A specific instance of this situation, which
is being extensively explored currently, concerns the properties of
ultra-cold, optically trapped atoms.  Of interest is how the
superfluid-insulator transition is modified by the inhomogeneity, and,
indeed, the extent to which a sharp transition survives at all.  This
paper explores a classical analog of these systems, the Blume-Capel
model with a spatially varying single ion anisotropy and/or
temperature gradient.  We present results both for the nature of the
critical properties and for the validity of the ``local density
approximation" which is often used to model the inhomogeneous case.
We compare situations when the underlying uniform transition is first
and second order.
\end{abstract}


\maketitle

\section*{Introduction}

The realization of superfluid and Mott insulator transitions in
optically trapped atoms \cite{greiner01,greiner02,stoferle04} has led
to an examination of the nature of phase transitions in the presence
of a spatially varying potential.  For example, it was found that when
a confining potential is added to the Bose-Hubbard Hamiltonian, the
variation of density across the sample results in a coexistence of
superfluid and Mott insulator regions \cite{batrouni02,batrouni08}.
As a consequence, critical phenomena which occur in the uniform case,
when the entire system collectively makes a transition from one phase
to another, are smeared.  The density no longer exhibits a singularity
as a function of chemical potential, as occurs in the translationally
invariant case \cite{fisher89,batrouni90}.  Measures of ``local
quantum criticality'' can be defined to help draw out residual signals
of the global phase transition \cite{rigol03}.

These conclusions have been drawn from direct examination of the
inhomogeneous model, but have also been inferred from studies of the
translationally invariant model combined with the ``local density
approximation" (LDA) \cite{bergkvist04}.  Specifically, the LDA
assumes that the properties of the confined system at a particular
spatial location are identical to those of the unconfined system with
a uniform potential taking the same value as the local potential at
that location.  Various checks have been made, for example by
comparing the LDA results using quantum monte carlo (QMC) of a
collection of uniform systems, with QMC simulations of a lattice with
a real trap \cite{bergkvist04,batrouni08}.

This LDA approximation is of course in direct analogy with that
commonly used in density functional theory \cite{lundqvist83}, where
the exact exchange-correlation potential present at a particular
position ${\bf r}$, in a system where the electron density varies
spatially, is replaced by the exchange correlation energy of the
uniform electron gas at the same constant density as that present at
${\bf r}$.  It is known that this approximation yields very good
results in a number of contexts, especially when the electron-electron
interactions are of weak to intermediate strength.  On the other hand,
when the coupling is stronger, and phenomena like magnetism and Mott
transitions occur, the LDA is less accurate \cite{kotliar06}.

In this paper, we examine the nature of phase transitions in spatially
inhomogeneous systems, and the validity of the LDA, within a more
simple classical context.  Previous work in this area includes studies
of Ising transitions in systems with a temperature gradient where the
nature of the interface between ferromagnetic regions adjacent to the
``cold side'' ($T<T_c$) of a sample and paramagnetic regions next to
the ``hot side" ($T>T_c$) has been explored
\cite{Boissin91,Batrouni92,Platini07,Hansen92}.

\section*{Model and Calculational Approach}

A classical model which can be constructed to have a spatially varying
density similar to that in optically trapped atom systems is the
Blume-Capel model \cite{Blume66,Capel66} with a site dependent
single-ion anisotropy,
\begin{eqnarray}
E = -J \sum_{\langle {\bf ij} \rangle} S_{\bf i} S_{\bf j}
   +  \sum_{\bf i} \Delta_{\bf i} S_{\bf i}^2 \,\,.
\end{eqnarray}
Here $S_{\bf i}$ is a discrete classical variable which can take on
three values, $S_{\bf i}= 0, \pm 1$.  A coupling $J$ is present
between near-neighbor spins which we choose to be positive
(ferromagnetic).  We consider a square lattice of linear size $L$.
That is, ${\bf i}=(i_x,i_y)$ with $1 \leq i_x,i_y \leq L$.  The value
$S_{\bf i}=0$ can be thought of as corresponding to a vacancy, while
$S_{\bf i}=\pm 1$ is an Ising spin, a collection of which can order
ferromagnetically if the ratio of $J$ to temperature $T$ is
sufficiently large. $\Delta$ is the single-ion anisotropy parameter, and
controls the density of $S_{\bf i}=0$ spins. 

The Blume-Capel model was originally introduced by
Blume~\cite{Blume66} and Capel~\cite{Capel66}, separately, to study
first-order magnetic transitions.  It was later generalized to the
Blume-Emery-Griffiths model (BEG) \cite{BEGriffiths}, which
incorporates an additional biquadratic interaction $K \sum_{\langle
  {\bf ij} \rangle} S_{\bf i}^2 S_{\bf j}^2$.  Since their initial
fomulation, the Blume-Capel and BEG models have been extensively used
to study the phase separation of He$^{3}$-He$^{4}$
mixtures~\cite{BEGriffiths} and various other systems that exhibit
tricritical behavior, such as multicomponent
fluids~\cite{Lajzerowicz75} and semiconductor alloys~\cite{Newman83}.
Recent works have used the Blume-Capel model to study ferromagnetic
thin films using an alternating single-ion
anisotropy~\cite{Zahraouy04}, and the dynamics of rough
surfaces~\cite{Brito07}.

Our computational method is standard Metropolis Monte Carlo.  Each
spin of the lattice is visited and a change from the current spin
value to one of the two other possibilities is suggested.  This change
is accepted or rejected with the Metropolis prescription.  To ensure
equilibration, a large number of sweeps of all the spins in the
lattice is performed prior to making measurements.  Unless otherwise
noted, the statistical errors in our results are smaller than the
symbol size.
The lattices studied in the manuscript are small enough that it is 
not necessary to emply more powerful cluster algorithms 
such as those developed by Swendsen and Wang \cite{swendsen87}.

For uniform systems, an accurate determination of the critical point
can be obtained from computing the second moment of the magnetization
\cite{landau00},
\begin{eqnarray}
\langle \, M^2 \, \rangle(T,L) 
= \frac{1}{L^4}\langle \,\,  (\sum_{\bf i} S_{\bf i} )^2 
\,\, \rangle \,\,.
\end{eqnarray}
Near the critical temperature, $T_c$, the following finite size
scaling expression holds,
\begin{eqnarray}
\langle M^2 \rangle(T,L) = L^{-2\beta/\nu} f[L^{1/\nu} (T_c-T)] \,\, .
\end{eqnarray}
Here $\beta$ ($\nu$) are the critical exponents governing how the
magnetization (correlation length) vanishes (diverges) as $T
\rightarrow T_c$ in the thermodynamic limit.  Eq.~3 implies that
plots of $L^{2\beta/\nu} \, \langle M^2 \rangle$ for different lattice
sizes $L$ cross at $T=T_c$, providing a method to locate the critical
temperature.

The physics of the Blume-Capel model with uniform $\Delta_{\bf i} =
\Delta$ is well understood. When $\Delta \rightarrow -\infty$,
vacancies ($S_{\bf i}=0$) are energetically very unfavorable.  The
system reduces to the Ising model and there is, on a square lattice, a
second order magnetic phase transition at $T_c = 2.269 J$. We can also
deduce the critical coupling at zero temperature.  The energy of the
fully polarized ferromagnetic state (all $S_{\bf i}=+1$) is $E_{\rm
  ferro} = (-2 J + \Delta) L^2$. The energy of the empty state (all
$S_{\bf i}=0$) is $E_{\rm vacuum}=0$. The ferromagnetic phase is
favored up until $\Delta > 2 J$.  Thus the phase diagram in the
$(T/J,\Delta/J)$ plane consists of a ferromagnetic region at low $T/J$
and low $\Delta/J$ bounded by the lines $T/J=2.269$ and
$\Delta/J=2$. As $\Delta$ increases from $\Delta=-\infty$ the extra
entropy of vacancies reduces $T_c$ until the Ising limit boundary
bends over to contact the $T=0$ critical point.

\begin{figure}[t]
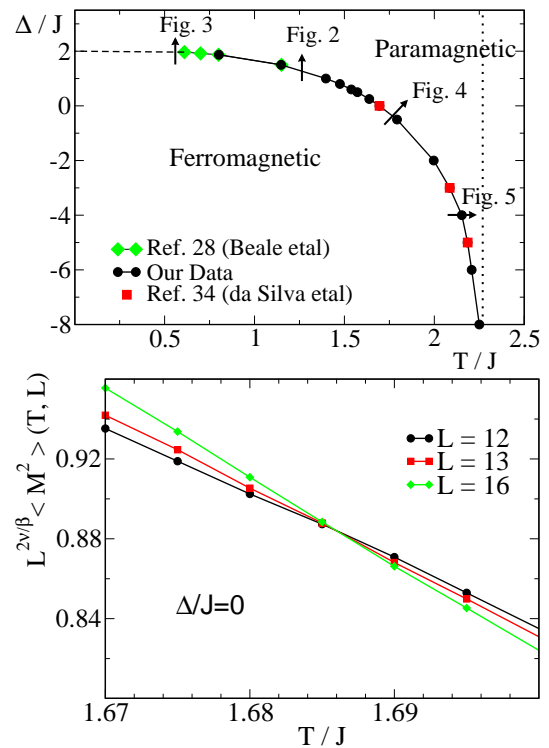

\centerline{\epsfig{figure=Fig1a.eps,width=7.0cm,angle=0,clip}}
\centerline{\epsfig{figure=Fig1b.eps,width=7.0cm,angle=0,clip}}
\caption{(color online) \underbar{Top:} Phase diagram of the Blume
  Capel model at uniform $\Delta_{\bf i}=\Delta$.  Second (first)
  order phase transitions are indicated by the solid (dashed) lines.
  The dotted line is the Ising limit.  The four diamonds depict values
  from Ref.~\onlinecite{PDBeale86}.  The squares are taken from
  Ref.~\onlinecite{Silva02}.  The rest of boundary was obtained using
  our code and the finite size scaling analysis of Eq.~3.  The arrows
  denote the trajectories used in the simulations of the inhomogeneous
  system, and correspond to Figs.~2,3,4,5 as indicated.  See text.
  \underbar{Bottom:} A representative finite-size scaling analysis is
  shown.  Here $\Delta=0$.  The critical temperature $T_c$ is
  determined by the position of the universal crossing of the scaled
  second moment of magnetization for different linear lattice sizes.
}
\label{fig:PhaseDiagram}
\end{figure}

The phase boundary for uniform $J,\Delta$ has been obtained by a
number of methods, including Monte Carlo
simulations~\cite{Wang87,Care93,Deserno97}, finite-size
scaling~\cite{PDBeale86,Plascak98}, renormalization group
methods~\cite{Berker76,Burkhardt76}, and series
expansions~\cite{Saul74}.  From these studies it is
known\cite{Plascak98} that there is a tricritical point along the
phase boundary at $(T/J,\Delta/J)=(0.609(4),1.965(5))$.  At low
temperatures in the vicinity of the $T=0$ critical point
$(T/J,\Delta/J)=(0,2)$ the magnetization jumps discontinuously upon
leaving the ferromagnetic phase.  Beyond the tricritical point, the
transition becomes continuous (second order).  The phase boundary for
this model is shown in Figure~\ref{fig:PhaseDiagram}(top), where the
values for $T_c$ were obtained through the analysis of Eqs.~2,3 using
our code and from Ref.~\onlinecite{PDBeale86} and
\onlinecite{Silva02}.  Figure~\ref{fig:PhaseDiagram}(bottom) shows a
representative finite size scaling crossing for $\Delta/J=0$.
Table~\ref{tab:criticaltemp} provides the locations of $T_c$ for
various values of $\Delta$.

\begin{table*}[htpb]
\centering
\begin{tabular}{cccc}
\hline
\hline
 $ \hskip0.2in \Delta/J$ \hskip0.2in 
&  $k_B T_c/J$ \ \ \  
& \ \ \ $k_B T_c/J$  
& \ \ \ $k_B T_c/J$  
\ \ \\
     &   (this work)\ \ \  
& \ \ \ (Ref.~\onlinecite{PDBeale86}) \ \ 
& \ \ \ (Ref.~\onlinecite{Plascak98}) \ \ \\
\hline
-8 & 2.250(4) &  & \\
-4 &  2.153(3)  &  & \\
-0.5 &  1.794(3)  &  & 1.794(7) \\
0 & 1.686(2) & 1.695  & 1.681(5) \\
1 & 1.397(1) & 1.398  & \\
1.87 & 0.802(2) & 0.800  & \\
\hline
\hline
\end{tabular}
\caption{Table of the critical temperatures for various values of
$\Delta /J$ that were found using our code and finite-size scaling
technique, in comparison with those from Refs.~\onlinecite{PDBeale86}
and \onlinecite{Plascak98}.  The value at $\Delta/J=-8$, where
vacancies are strongly suppressed, is close to the
$T_c/J = 2.269$ of the two dimensional Ising model, as expected.}
\label{tab:criticaltemp}
\end{table*}

Having reviewed and reproduced some of the features of the
transitionally invariant Blume-Capel model, 
we now turn to the subject of this paper, the inhomogeneous case. We
choose three models of spatial inhomogeneity.  In the first two
we introduce a linear variation of either the 
single-ion anisotropy or the temperature, keeping the other parameters
fixed,
\begin{eqnarray}
\Delta({\bf i}) &=& \Delta_0 + 
\frac{\Delta_1-\Delta_0}{L_x}\,\,i_x, \ \ \ \ T = {\rm const}
\nonumber
\\
T({\bf i}) &=& T_0 + \frac{T_1-T_0}{L_x}\,\,i_x, \ \ \ \ \Delta = {\rm
const}
\end{eqnarray}
These correspond to vertical ($\Delta$ varying) and horizontal 
($T$ varying) cuts in the phase diagram.
In the third case we allow both temperature and single ion anisotropy
to change together,
\begin{eqnarray}
   \Delta({\bf i}) &=& \Delta_0 +
  \frac{\Delta_1-\Delta_0}{L_x}\,\,i_x,
 \nonumber
 \\
 T({\bf i}) &=& m \Big( \, \Delta_1 +
\frac{\Delta_1-\Delta_0}{L_x}\,\,i_x\, \Big)
\end{eqnarray}
where $m$ determines the slope of $T({\bf i})$. 
This more general inhomogeneity allows us to 
follow paths in the $(T,\Delta)$ plane which are perpendicular
to the phase boundary in the intermediate coupling regime where
the boundary curves around from its low $T$ and large negative
$\Delta$ limits.  
Typically we will
be interested in cases where $\Delta_0, \Delta_1,T_0$ and $T_1$ are
chosen such that the lattice is ferromagnetic on the left side,
$i_x=1$, with very few vacancies, and then becomes paramagnetic for
$i_x = L_x$.  For simplicity, we have chosen a gradient only in one
spatial direction $x$, so the iso-contours of the single ion
anisotropy are vertical lines.  In $d=2$ ultracold trapped gases, the
iso-contours are typically circles around the trap center.  However,
we do not expect the results of our study to depend on the shape of
the boundary between phases, only on the existence of the boundary
itself\cite{wessel}.

We have imposed periodic boundary conditions (pbc) in both the $x$ and
$y$ directions.  Besides reducing finite size effects, the use of pbc
avoids having edge sites with a smaller number of neighbors than in
the bulk, a situation which would make the connection with the LDA
less simple.  However, there is one slightly tricky issue with the
pbc.  The pbc links in the $y$ direction by construction connect sites
with the same $\Delta_{\bf i}$.  In the x-direction, the pbc's link
sites with vastly different single ion anisotropies: $\Delta_0$ and
$\Delta_1$.  To avoid this problem, the simulations were run on
lattices with linear size $2L_x+1$ in the x-direction with $\Delta_{\bf
  i}$ symmetric across the center of the lattice.  In effect, a second
copy of the lattice is connected to the $x=1$ boundary of the first,
and the values of $\Delta_{\bf i}$ increase linearly back up to
$\Delta_1$ at which point the pbc connection is established.  A final
point about the geometry is that when we explore finite sizes effects
we will fix $L_y=50$ and increase $L_x$ at constant $(\Delta_1 -
\Delta_0)$.  This is done because the $x$ direction is the one along
which the gradient is established and so increasing $L_x$ allows us to
explore the limit where the anisotropy gradient becomes weaker and weaker.
Each of the cases will be used in regions where the respective
gradient is approximately perpendicular to the phase boundary, as
shown in figure~\ref{fig:PhaseDiagram}(top).  This ensures that the
critical region will be localized to a small area of the lattice.

We present results for the ``linear structure factor," which we define
as,
\begin{eqnarray}
{\bf \cal S}(i_x) &=& 
\langle \, \frac{1}{L_y^2} 
( \sum_{i_y}  S_{(i_x,i_y)} )^2  \, \rangle 
\nonumber \\
&=& \frac{1}{L_y}  \sum_{l_y}  \langle \,
S_{(i_x,1)} S_{(i_x,1+l_y)} \, \rangle \,\, .
\end{eqnarray}
${\bf \cal S}(i_x)$ sums up the spin-spin correlations for all
separations $l_y=1,2,\cdots L$ with a given $i_x$.  The pairs of sites
in ${\bf \cal S}(i_x)$ therefore all have the same value of
$\Delta_{\bf i}$.  This is a convenient (indeed essential) choice in
order to make meaningful comparisons with the LDA which employs
lattices of constant $\Delta$.  In this way, ${\bf \cal S}(i_x)$ is
the natural generalization of the mean square magnetization (Eq.~2)
used in the translationally invariant case.

We will also compare the energy for an inhomogeneous lattice with that
obtained by the LDA.  Similar considerations apply here as with the
linear structure factor; we would like to compare observables for sets
of sites with the same value of $\Delta_{\bf i}$.  However, the energy
involves links (in the x-direction) which connect sites with different
$\Delta_{\bf i}$.  For this reason we will present results for the
energy associated with bonds only in the $y$ direction,
\begin{eqnarray}
E_y(i_x) &&= -\frac{J}{L_y} \sum_{i_y}  S_{(i_x,i_y)} S_{(ix,i_y+1)}
\nonumber \\
&&+ \frac{1}{L_y} \sum_{i_y}  \Delta_{(i_x,i_y)} S_{(i_x,i_y)}^2  \,\,.
\end{eqnarray}

\section*{\bf Results}

\begin{figure}[t]
\centerline{\epsfig{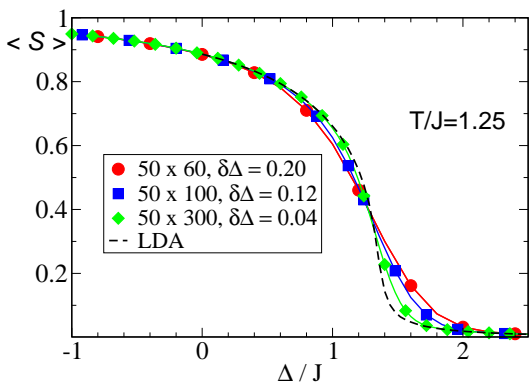}}
\caption{(color online) The full lines depict the linear structure
  factor ${\cal S}$ versus $\Delta({\bf i})$ at $T/J=1.25$ for several
  different gradients in the single-ion potential.  For all full lines
  $\Delta_0/J=-8.00$ and $\Delta_1/J=4.00$, but as $L_x$ increases the
  gradient $\delta \Delta = (\Delta_1 - \Delta_0) / L_x$ softens.  The
  LDA result is the dashed curve, and is quantitively correct except
  in the transition region.  As expected, the accuracy of the LDA
  improves as the gradient of the inhomogeneity decreases.  
}
\label{fig:2ndOrderDG}
\end{figure}

Figure~\ref{fig:2ndOrderDG} shows, for fixed $T/J=1.25$, the linear
structure factor ${\cal S}$ as a function of $\Delta/J$.  More
precisely, ${\cal S}(i_x)$ is computed at different values of $i_x$
for a system with a gradient $\delta \Delta= (\Delta_1-\Delta_0)/L_x$
with $\Delta_0/J=-8.00$ and $\Delta_1/J=4.00$.  The value of ${\cal
  S}(i_x)$ is plotted against the corresponding value of
$\Delta_{(i_x,i_y)}$ on the horizontal axis.  Since the relation
between $\Delta_{(i_x,i_y)}$ and $i_x$ is linear (Eq.~4), the
horizontal axis can equivalently be viewed as labeling the spatial
position as one sweeps across the inhomogeneous lattice.  At the same
time, the LDA values are obtained by simulating uniform systems at
a range of $\Delta$ values corresponding to the vertical trajectory
marked ``Fig. 2" in the phase diagram of the uniform system,
Fig.~\ref{fig:PhaseDiagram}(top).  This trajectory crosses the
ferromagnetic to paramagnetic phase boundary at $\Delta = 1.300(3)$ in a
second order transition.  We see that the LDA predicts the behavior of
${\cal S}$ in a qualitatively correct fashion over the entire range of
$\Delta$, and is quantitively accurate except in the vicinity of the
critical region where the lattice inhomogeneity blurs the transition.
This is the same basic result as found for optically trapped atom
systems \cite{batrouni02,batrouni08}.  However, we are able in this
simple classical model to compare more precisely the LDA with the
inhomogeneous case.  In particular, Fig.~\ref{fig:2ndOrderDG} shows
the improved accuracy of the LDA as the gradient of the inhomogeneity
becomes smaller, something which has not yet been done in the quantum
case.

\begin{figure}[t]
\centerline{\epsfig{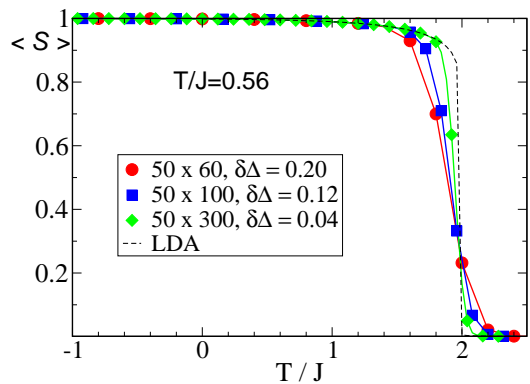}}
\caption{(color online) ${\cal S}$ versus $\Delta({\bf i})$ at
  $T/J=0.56$, a temperature below the tricritical point.  Data are
  shown for several different values of the single-ion potential
  gradient (full curves).  While the LDA predicts a first order phase
  transition (dashed curve), as is expected at this temperature,
  the phase transitions of the inhomogenous systems are less abrupt.
  As the variations in the single-ion potential decrease, or one moves
  far from the transition region, ${\cal S}$ converges to the results
  obtained by the LDA.  
}
\label{fig:1stOrderDG}
\end{figure}

Fig.~\ref{fig:1stOrderDG} shows a similar set of data but for
$T/J=0.56$ which corresponds to the trajectory labeled ``Fig. 3" in
Fig.~1 and crosses the ferromagnetic-paramagnetic phase boundary in a
first order transition at $\Delta/J=1.979(2)$.  Again, the LDA is
qualitatively correct.  Comparing Figs.~\ref{fig:2ndOrderDG} and
\ref{fig:1stOrderDG}, it appears that the LDA has larger quantitative
errors in the vicinity of the transition region in the first order
case, but that these errors extend less far away from the transition
region.  This result seems reasonable: a smoothly varying potential
does not exhibit the very abrupt discontinuity in the LDA
results, but because the first order transition
region is narrower, the region where LDA fails significantly is
less wide.
It is notable that curves of the linear structure factor for different
values of the gradient cross at roughly a single point.  

\begin{figure}[t]
\centerline{\epsfig{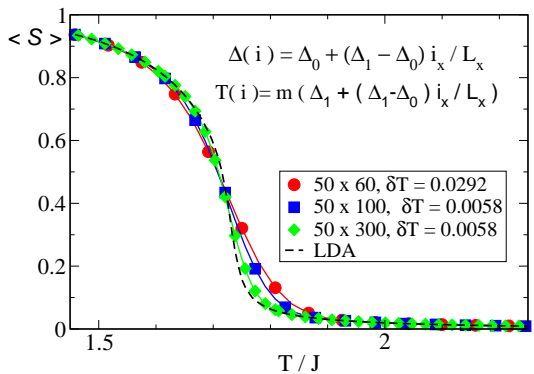}}
\caption{(color online) The linear structure factor ${\cal S}$ versus
  $T(\Delta({\bf i}))$ for the inhomogenous case with both a spatially
  varying single-ion potential and temperature gradient.  The choice 
$m=0.1458$ in Eq.~5 makes this trajectory cross perpendicular to the phase 
boundary.}
\label{fig:MixedGradient}
\end{figure}

\begin{figure}[t]
\centerline{\epsfig{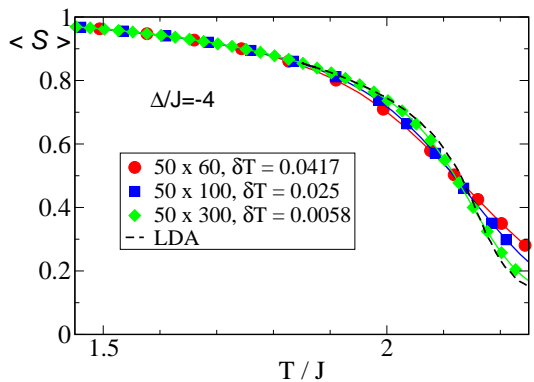}}
\caption{(color online) The linear structure factor ${\cal S}$ versus
  $T({\bf i})$ at $\Delta/J=-4$ for several different values of
  temperature gradients.}
\label{fig:ComparisonSTG}
\end{figure}

Our final two results for the linear structure factor are given in
Figs.~\ref{fig:MixedGradient} and \ref{fig:ComparisonSTG}, and show
cuts across the phase boundary in which both temperature and the
single ion anisotropy are simultaneously evolving.  The trajectories
are labeled Figs.~\ref{fig:MixedGradient} and \ref{fig:ComparisonSTG},
in Fig.~\ref{fig:PhaseDiagram}.  The physics of our model does not
depend independently on $T, \Delta_{\bf i},$ and $J$, but only on the
ratios $\Delta_{\bf i}/T$ and $J/T$.  In Fig.~\ref{fig:2ndOrderDG}
only the first of these ratios is changing, while in
Figs.~\ref{fig:MixedGradient} and \ref{fig:ComparisonSTG} both ratios
are evolving as we traverse the lattice.  Since both of these cuts
traverse the phase boundary in the second order region, the results
for the LDA resemble those of Fig.~\ref{fig:2ndOrderDG}.  This
emphasizes that the question of the accuracy of the LDA does not
appear crucially to depend on which parameters in the energy (or the
temperature) are varying.

\begin{figure}[t]
\centerline{\epsfig{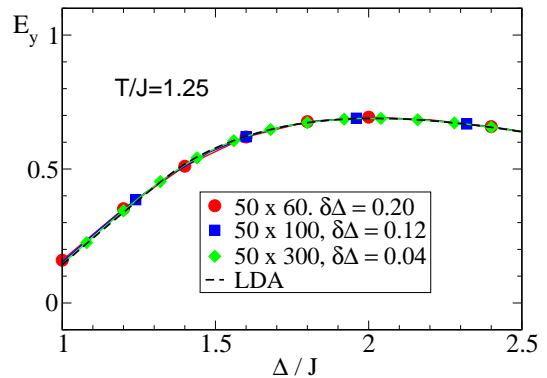}}
\caption{(color online) Comparison of the LDA prediction for the
  energy with the energy of an inhomogeneous system.  Here we have
  fixed $T/J=1.25$ and are changing $\Delta_{\bf i}$ across the
  lattice.  This is the trajectory labeled Fig.~\ref{fig:2ndOrderDG}
  in Fig.~\ref{fig:PhaseDiagram}.  The LDA energy is remarkably
  accurate even in the transition region.  }
\label{fig:ComparisonEnergy2ndDG}
\end{figure}

We now turn to a comparison of the energy of the inhomogeneous system
with that of the LDA.  Fig.~\ref{fig:ComparisonEnergy2ndDG} shows the
same cut at constant $T/J=1.25$ as in Fig.~\ref{fig:2ndOrderDG}.
Remarkably, the energy is given very accurately by the LDA throughout
the inhomogeneous lattice, even through the transition region where
the linear structure factor differed markedly.  This result may appear
surprising in that the first piece of $E_y$ in Eq.~5 is the near
neighbor spin correlation in the y-direction, which is also one of the
ingredients of the linear structure factor ${\cal S}$.  That the LDA
value for $E_y$ is so accurate suggests that the failure of the LDA in
the transition region is dominated by its mis-estimate of the {\it
  long-range} correlations, while the short range-ones are correctly
identified.  Indeed, this result might be expected since it is the
long-range pieces of ${\cal S}$ whose behavior is crucial to the
occurrence of a second order transition.

\begin{figure}[t]
\centerline{\epsfig{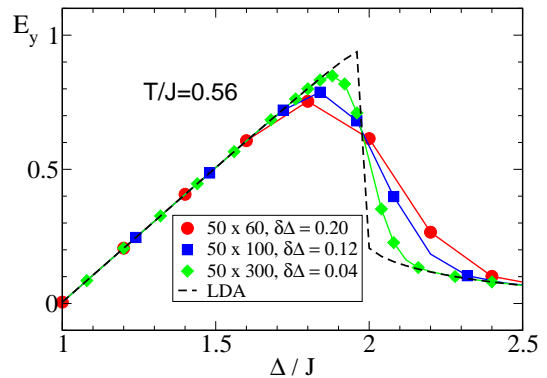}}
\caption{(color online) Comparison of the LDA prediction for the
  energy with the energy of an inhomogeneous system.  Here we have
  fixed $T/J=0.56$ and are changing $\Delta_{\bf i}$ across the
  lattice.  This is the trajectory labeled Fig.~\ref{fig:1stOrderDG}
  in Fig.~\ref{fig:PhaseDiagram}.  Near the transition the LDA energy
  differs markedly from that of an inhomogeneous lattice.  }
\label{fig:EnergyComparisonDG}
\end{figure}

Finally, Fig.~\ref{fig:EnergyComparisonDG} shows the same cut at
constant $T/J=0.56$ as in Fig.~\ref{fig:1stOrderDG}.  Here, when the
underlying homogeneous transition is first order, we see that even the
energy is badly estimated by the LDA.  The local energy has a
universal crossing, corresponding to the transition value of
$\Delta/J$, similar to that of the linear structure factor ${\cal S}$.

\section*{Conclusions}

The ``Local Density Approximation" is a commonly employed method to
understand the phase transitions of ultracold, optically trapped atoms
which experience a spatially varying confining potential.  In this
paper we have explored the validity of the LDA in the simpler,
classical, Blume-Capel model to which we have applied a gradient in
the temperature and/or the single ion anisotropy.

Our basic conclusion is that the LDA performs well quantitatively in
regions that are not close to where the state of the system is making
a transition between the allowed phases, in our case ferromagnetic and
paramagnetic.  That is, the values of the local structure factor and
energy predicted by the LDA match those of a direct simulation of the
inhomogeneous system except in the transition zone.  This is similar
to the conclusions drawn in the optical lattice case
\cite{batrouni02}.  However, because our model is classical as opposed
to quantum mechanical, we can explore the validity of the
LDA in greater detail, including the systematic improvement in the
accuracy of the LDA with inhomogeneous systems which have smaller
gradients.

An especially interesting feature of the Blume-Capel model is the
presence of a tricritical point on the phase boundary.  This allows us
to compare the validity of the LDA for first and second order
transitions in the same model.  Our conclusion is that a quantity like
the linear structure factor which samples long range correlations is
more badly estimated by the LDA in the transition region of a first
order phase change, but that the width of the region over which the
LDA is inaccurate is more narrow.  Overall, the accuracy of the LDA in
the two cases is not so dramatically different.  On the other hand,
the predictive accuracy of the LDA for the energy, which samples just
short range correlations, is very different for the first and second
order situations.  In the second order case, the LDA energy is
quantitativly correct even through the transition region, while in the
first order case the energy is rather badly mis-estimated.

This work was supported by CNRS (France) PICS 18796 and ARO Award
W911NF0710576 with funds from the DARPA OLE Program.  We acknowledge
useful input from T.~Hollies.


\begin{thebibliography}{10}

\bibitem{greiner01}
M. Greiner, I. Bloch, O. Mandel, T.W. H\"ansch, and T. Esslinger,
Phys. Rev. Lett. {\bf 87}, 160405 (2001).

\bibitem{greiner02}
M. Greiner, O. Mandel, T. Esslinger, T. W.
H\"ansch and I. Bloch, Nature {\bf 415}, 39 (2002).

\bibitem{stoferle04}
T. St\"oferle, H. Moritz, C. Schori, M. K\"ohl, and T. Esslinger,
Phys. Rev. Lett. {\bf 92}, 130403 (2004).

\bibitem{batrouni02}
G.G. Batrouni, V. Rousseau,
R.T.~Scalettar, M.~Rigol, A.~Muramatsu, P.J.H. Denteneer, and M. Troyer,
Phys.~Rev.~Lett.~{\bf 89}, 117203 (2002).

\bibitem{batrouni08}
G.G. Batrouni, H.R. Krishnamurthy, K. Mahmud, V.G. Rousseau,
and R.T. Scalettar,
Phys.~Rev.~{\bf A}, to appear.

\bibitem{fisher89}
 M.P.A. Fisher, P.B. Weichman, G. Grinstein, and D.S. Fisher, Phys. Rev. {\bf B40}, 546 (1989).

\bibitem{batrouni90}
G.G.~Batrouni, R.T.~Scalettar, and G.T.~Zimanyi,
Phys.~Rev.~Lett. {\bf 65}, 1765 (1990).

\bibitem{rigol03}
M.~Rigol, A.~Muramatsu, G.G. Batrouni, and R.T.~Scalettar,
Phys. Rev. Lett. {\bf 91}, 130403 (2003).

\bibitem{bergkvist04}
S. Bergkvist, P. Henelius, and A. Rosengren, 
Phys. Rev.  {\bf A70}, 053601 (2004).

\bibitem{lundqvist83}
{\it Theory of the Inhomogeneous Electron
Gas}, edited by S. Lundqvist and S. H. March (Plenum, New
York, 1983).

\bibitem{kotliar06}
G. Kotliar, S. Y. Savrasov, K. Haule, V. S.
Oudovenko, O. Parcollet,
and C. A. Marianetti, Rev. Mod. Phys. {\bf 78}, 865 (2006).

\bibitem{Boissin91}
N. Boissin and H.J. Herrmann, J. Phys. A:Math Gen. {\bf 24}, L43-L45 (1991).

\bibitem{Batrouni92}
G.G. Batrouni and A. Hansen, J. Phys. A:Math Gen. {\bf 25}, L1059-L1064 (1992).

\bibitem{Platini07}
Gradients in the quantum Ising chain are explored in:
T. Platini, D. Karevski, and L. Turban,
J. Phys. A:  Math. Theor. {\bf 40}, 1467 (2007).
The quantum Ising chain maps onto the 2d classical Ising model.
This paper also includes a further review of the literature concerning
inhomogeneities.

\bibitem{Hansen92}
A. Hansen and D. Stauffer,
Physica {\bf A189}, 611 (1992).

\bibitem{Blume66}
M.~Blume, Phys. Rev. {\bf 141}, 517 (1966)

\bibitem{Capel66}
 H.W.~Capel, Physica {\bf 32}, 966 (1966).

\bibitem{BEGriffiths}
M.~Blume, V.J.~Emery, and R.B.~Griffiths, Phys. Rev. {\bf A4}, 1071 (1971).

\bibitem{Lajzerowicz75}
J. Lajzerowicz and J. Sivardiere, Phys. Rev. {\bf A11}, 2079 (1975).

\bibitem{Newman83}
K.E.~Newman and J.D. Dow, Phys. Rev. {\bf B27}, 7495 (1983).

\bibitem{Zahraouy04}
H. Ez-Zahraouy and A. Kassou-Ou-Ali, Phys. Rev. {\bf B69}, 064415 (2004).

\bibitem{Brito07}
A. Brito, J. A. Redinz, and J. A. Plascak, Phys. Rev. {\bf E75}, 046106 (2007).

\bibitem{swendsen87}
R.H. Swendsen and J-S. Wang,
Phys. Rev. Lett. {\bf 58}, 86 (1987).

\bibitem{landau00}
``A Guide to Monte Carlo Simulations in Statistical
Physics," D.P. Landau and K. Binder, Cambridge University Press (2000).

\bibitem{Wang87}
Y. Wang, F. Lee, and J.D. Kimel, Phys. Rev. {\bf B36}, 8945 (1987).

\bibitem{Care93} 
C. M. Care, J. Phys. {\bf A26}, 1481 (1993).

\bibitem{Deserno97}
M. Deserno, Phys. Rev. {\bf E56}, 5204 (1997).

\bibitem{PDBeale86}
P.D. Beale, Phys. Rev. {\bf B33}, 1717 (1986).

\bibitem{Plascak98}
J.C. Xavier, F.C. Alcaraz, D.Pena Lara, and J.A. Plascak, Phys. Rev.
{\bf B57}, 11575 (1998).

\bibitem{Berker76}
A. N. Berker and M. Wortis, Phys. Rev. {\bf B14}, 4946 (1976).

\bibitem{Burkhardt76}
T. W. Burkhardt, Phys. Rev. {\bf B14}, 1196 (1976)

\bibitem{Saul74}
D. M. Saul, M. Wortis, and D. Stauffer, Phys. Rev. {\bf B9}, 4964
(1974).

\bibitem{wessel} S. Wessel, F. Alet, M. Troyer, and G. G. Batrouni,
  Phys. Rev. {\bf A70}, 053615 (2004).

\bibitem{Silva02}
Roberto da Silva, Nelson A. Alves, and J.R. Drugowich de Felicio,
Phys. Rev. {\bf E66}, 026130 (2002).

\bibitem{spielman08}
I. B. Spielman, W. D. Phillips, and J. V. Porto,
Phys. Rev. Lett. {\bf 100}, 120402 (2008).

\end{thebibliography}
\end{document}